\documentclass[journal=jacsat,manuscript=article]{achemso}

\usepackage{graphicx}
\usepackage{dcolumn}
\usepackage{bm}

\usepackage[utf8]{inputenc}
\usepackage[T1]{fontenc}
\usepackage{mathptmx}
\usepackage{etoolbox}

\usepackage{bm}
\usepackage{esvect}
\usepackage{amsmath}
\usepackage{braket}
\usepackage[version=3]{mhchem} 

\author{Yifei Geng}
  \affiliation{School of Electrical and Computer Engineering, Cornell University, Ithaca, New York 14853, USA.}
  \email{yg474@cornell.edu}

\author{Debdeep Jena}
	
\affiliation{School of Electrical and Computer Engineering, Cornell University, Ithaca, New York 14853, USA.}
\alsoaffiliation{Department of Materials Science and Engineering, Cornell University, Ithaca, New York 14853, USA.}
	
\author{Gregory D. Fuchs}
\affiliation{School of Applied and Engineering Physics, Cornell University, Ithaca, New York 14853, USA.}

 \author{Warren R. Zipfel}
\affiliation{Meinig School of Biomedical Engineering, Cornell University, Ithaca, New York 14853, USA.}
 
\author{Farhan Rana}
\affiliation{School of Electrical and Computer Engineering, Cornell University, Ithaca, New York 14853, USA.}

\title[An \textsf{achemso} demo]
  {Optical Dipole Structure and Orientation of GaN Defect Single-Photon Emitters}


\begin{document}



\begin{abstract}
  GaN has recently been shown to host bright, photostable, defect single photon emitters in the 600-700 nm wavelength range that are promising for quantum applications. The nature and origin of these defect emitters remain elusive. In this work, we study the optical dipole structures and orientations of these defect emitters using the defocused imaging technique. In this technique, the far-field radiation pattern of an emitter in the Fourier plane is imaged to obtain information about the structure of the optical dipole moment and its orientation in 3D. Our experimental results, backed by numerical simulations, show that these defect emitters in GaN exhibit a single dipole moment that is oriented almost perpendicular to the wurtzite crystal c-axis. Data collected from many different emitters shows that the angular orientation of the dipole moment in the plane perpendicular to the c-axis exhibits a distribution that shows peaks centered at the angles corresponding to the nearest Ga-N bonds and also at the angles corresponding to the nearest Ga-Ga (or N-N) directions. Moreover, the in-plane angular distribution shows little difference among defect emitters with different emission wavelengths in the 600-700 nm range. Our work sheds light on the nature and origin of these GaN defect emitters. 
\end{abstract}

\section{Introduction}
Single photon emitters (SPEs) play an important role in quantum computing and quantum communication technologies~\cite{aharonovich2016solid}.
SPEs with high brightness, high photon indistinguishability, narrow emission spectrum, small blinking effects and integration compatibility with technologically important material platforms are desired for quantum applications. A family of defect based SPEs in GaN in the 600-700 nm wavelength range have been reported to exhibit high brightness and sharp zero phonon lines (ZPL) even at room temperature~\cite{berhane2017bright,berhane2018photophysics}, making these defects very promising for single-photon applications. Recently, optically detected magnetic resonance (ODMR) was also reported in these GaN defect states~\cite{Luo23}. However, very little is known thus far about the origin of these defects. Point defects~\cite{geng2023dephasing} in the form of substitutional impurity atoms, or impurity-vacancy complexes, and defect states near dislocations or stacking faults~\cite{berhane2017bright,berhane2018photophysics} have been proposed as candidates for these SPEs in the literature. The identification of the nature and origin of these defect SPEs is an important first step in harnessing their potential for applications.

\begin{figure}
\includegraphics[width=0.7\textwidth]{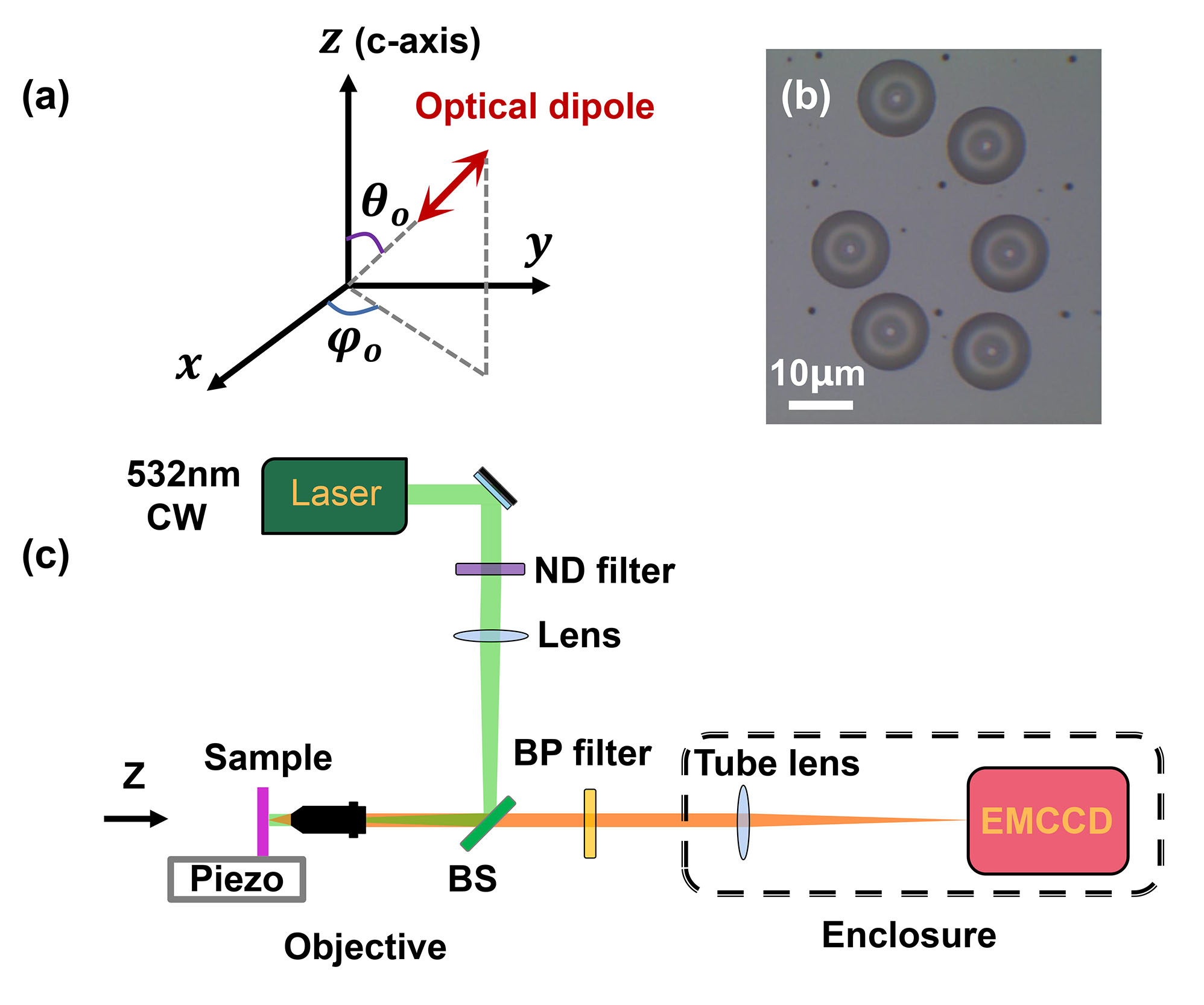}
\caption{\label{fig1} (a) The spatial orientation of a SPE optical dipole moment is specified by the angles $\varphi_{o}$ and $\theta_{o}$. (b) Six fabricated solid immersion lenses (hemispheres of 5 $\mu$m  diameter) are shown in the optical microscopic image. (c) Measurement setup for defocused imaging.}
\end{figure}

A topic of special interest in this context is the determination of the spatial orientation of the defect's optical dipole moment. The spatial orientation of the dipole moment, as shown in Fig.~\ref{fig1} (a), is specified by the angle $\theta_{o}$ from the z-axis (henceforth assumed to be parallel to the c-axis of the wurtzite crystal) and the angle $\varphi_{o}$ in the plane perpendicular to the c-axis. On one hand, the dipole orientation strongly affects the design and efficiency of optical structures (gratings, microcavities, waveguides, etc) needed to couple photons out of the defect, and on the other hand, the dipole orientation can say a lot about the nature of the defect and its environment. Few different experimental techniques have been demonstrated to determine the 3D orientation of the optical dipole moment of an emitter\cite{zhou2021experimental,6466354,dolan2014complete}. The simplest ones among these are based on light collection from the dipole and either record the image at the back focal plane of the objective~\cite{Lieb04} or record the emission pattern with an objective that is defocused with respect to the dipole~\cite{Jasny97,Bohmer03,patra2004image,gardini20153d}, and then use analytical models or numerical computations to determine the orientation of the dipole from the recorded image data. The accuracy of both these techniques is generally around $\pm$15 degrees but can be as good as $\pm$5 degrees depending on the photon collection efficiency of the measurement setup and the available signal to noise ratio. The defocused imaging technique has been previously used to determine the optical dipole orientation in colloidal quantum dots\cite{schuster2005defocused,patra2005defocused,lethiec2014polarimetry} and dye molecules~\cite{habuchi2011multi,su2016super}.

In this work, we use the defocused imaging technique to determine the optical dipole orientations of GaN defect emitters. Our results show that the GaN defect emitters have a single dipole moment (and not an incoherent mixture of more than one dipole moments~\cite{patra2005defocused}) that is oriented nearly perpendicular to the c-axis. Data collected from many different emitters shows that the angular orientation of the dipole moment in the plane perpendicular to the c-axis exhibits a distribution that shows peaks centered at the angles corresponding to the nearest Ga-N bonds and also at the angles corresponding to the nearest Ga-Ga (or N-N) directions. Interestingly, the in-plane angular distribution shows little difference among defect emitters with different emission wavelengths in the 600-700 nm range. Our results are consistent with these emitters being point defects due to substitutional or interstitial impurities (or vacancy-impurity complexes). 

\section{Results and Discussion}

The GaN SPEs under investigation were hosted in 4 $\mu m$ thick HVPE grown semi-insulating GaN epitaxial layers on 430 $\mu m$ sapphire substrates. The surface normal direction is the crystal c-axis. The large refractive index contrast between air and GaN reduces, in many different ways, the accuracy with which the orientation of an optical dipole can be determined using defocused imaging. For one, it reduces the photon collection efficiency and the signal to noise ratio. Second, it causes optical standing waves to form in the epitaxial layer making computational determination of the dipole orientation difficult from the recorded data. Finally, the inability to collect photons emitted at large angles from the surface normal direction (as a result of total internal reflection) also degrades the accuracy. To overcome this problem, solid immersion lenses were fabricated around each defect emitter in the form of hemispheres of diameters 5 $\mu$m by focused ion beam milling~\cite{geng2023dephasing}, as shown in Fig.~\ref{fig1}(b). An optical microscope setup, shown in Fig.~\ref{fig1}(c), incorporating wide-field laser illumination, was used for pumping the SPEs with 532 nm wavelength light. Emitted light was collected using an oil-immersion objective lens (NA=1.4), passed through a bandpass filter, and then directed at an EMCCD camera (Andor, Oxford Instruments). The sample was mounted on a piezostage. The objective was used to record the radiation pattern of a SPE by defocusing the objective with respect to the SPE. The defocusing was achieved by moving the piezostage about 1 $\mu$m away from the focal plane of the objective~\cite{Bohmer03,patra2004image,gardini20153d}. The EMCCD accumulated 20 frames with an electron multiplying (EM) gain of 500, and had an exposure time of 0.1 s for each frame. A home-built confocal scanning microscope with a Hanbury-Brown and Twiss interferometer was used to measure the emission spectra, polarization patterns, and the second order correlation functions ($g^{(2)}$) of SPEs\cite{geng2023dephasing}.

\begin{figure*}
\includegraphics[width=0.9\textwidth]{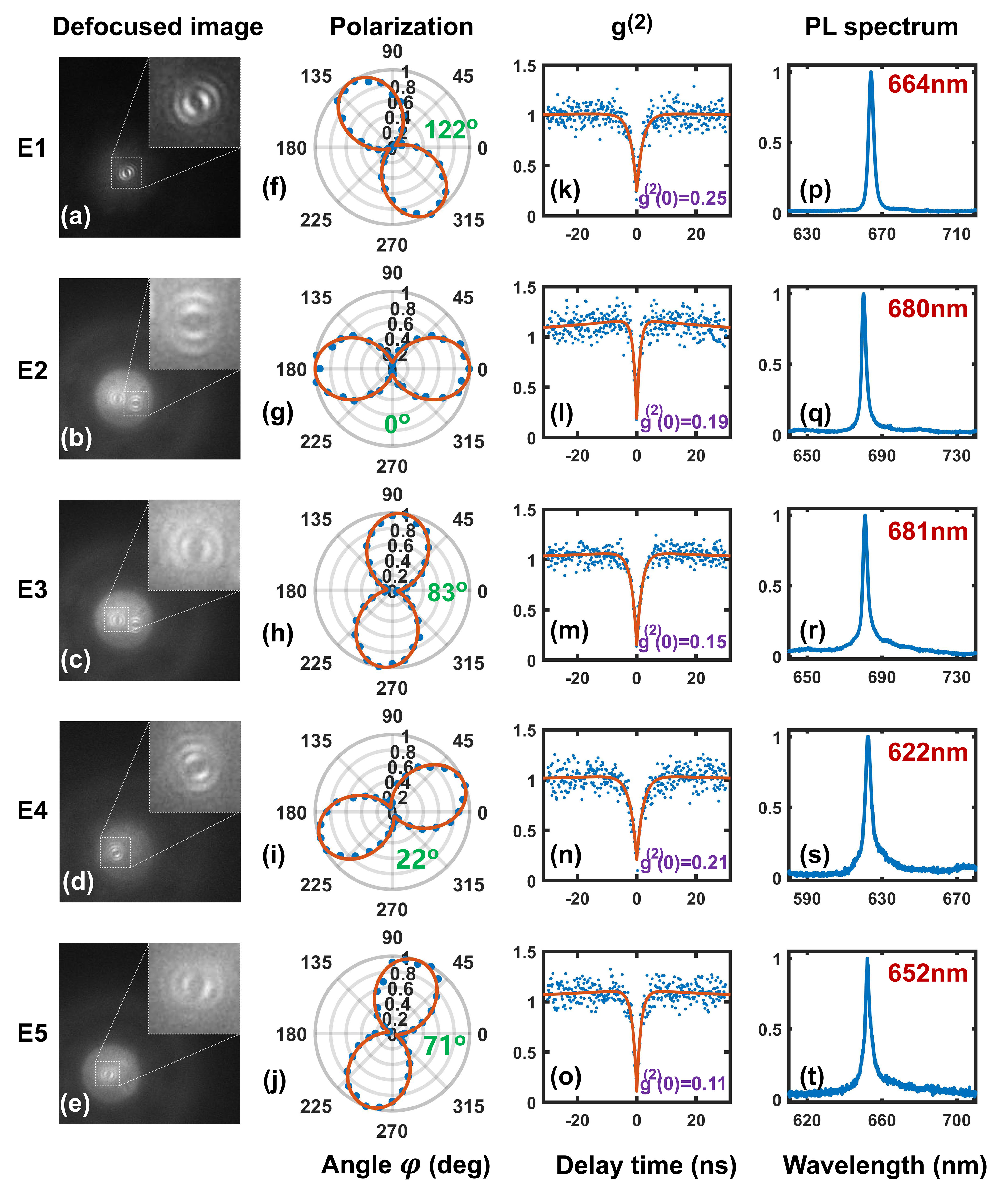}
\caption{\label{fig2} (a)-(e) Defocused images of light emission from five different SPEs, E1 through E5, integrated with solid immersion lenses are shown. Inserts show the zoomed-in radiation patterns of each SPE. (f)-(j) Measured polarization patterns of the emitted light are plotted as a function of the in-plane angle $\varphi$. $\varphi=0^{\circ}$ corresponds to the direction perpendicular to the wurtzite m-plane. The texts in green show the in-plane angles $\varphi_{o}$ of the dipoles as obtained by fitting data with the function $\cos^{2}(\varphi - \varphi_{o})$. (k)-(o) Measured second order correlation functions are plotted. (p)-(t) Measured photoluminescence (PL) spectra are plotted. All measurements were performed at room temperature.}
\end{figure*}

More than a hundred different SPEs were studied in this work. All measurements were carried out at room temperature. Fig.~\ref{fig2} shows representative data from five different emitters. Figs.~\ref{fig2}(a)-(e) show the radiation patterns (defocused images) of emitters E1 through E5 integrated with solid immersion lenses (emitter E2 and E3 belong to the same solid immersion lens).
All SPEs exhibit a two-lobe radiation pattern which, as shown later, is consistent with a single optical dipole oriented almost perpendicular to the crystal c-axis. 
Figs.~\ref{fig2}(f)-(j) show the corresponding polarization patterns. The polarization data in each case can be fitted with the function $\cos^{2}(\varphi - \varphi_{o})$, consistent with light polarized linearly in a direction $\varphi_{o}$ that is in agreement with the measured radiation pattern. $\varphi=0^{\circ}$ corresponds to the direction perpendicular to the wurtzite m-plane. The texts in green in Figs.~\ref{fig2}(f)-(j) show the in-plane angles $\varphi_{o}$ of the dipoles as obtained by fitting polarization patterns with the function $\cos^{2}(\varphi - \varphi_{o})$. 
Figs.~\ref{fig2}(k)-(o) show the second order correlation functions ($g^{(2)}$) measured using the time-tagged time-resolved (TTTR) mode of the correlator (MultiHarp150 from Picoquant) instrument. The small values (below 0.5) of $g^{(2)}(0)$ confirm that all defects are SPEs. The measured emission spectra are plotted in Figs.~\ref{fig2}(p)-(t). All SPEs exhibit sharp and strong zero phonon lines in the 600-700 nm wavelength range at room temperature. Although the emission wavelengths are different, all emitters in Figs.~\ref{fig2} show similar defocused radiation patterns and similar polarization patterns.

\begin{figure}
\includegraphics[width=0.6\textwidth]{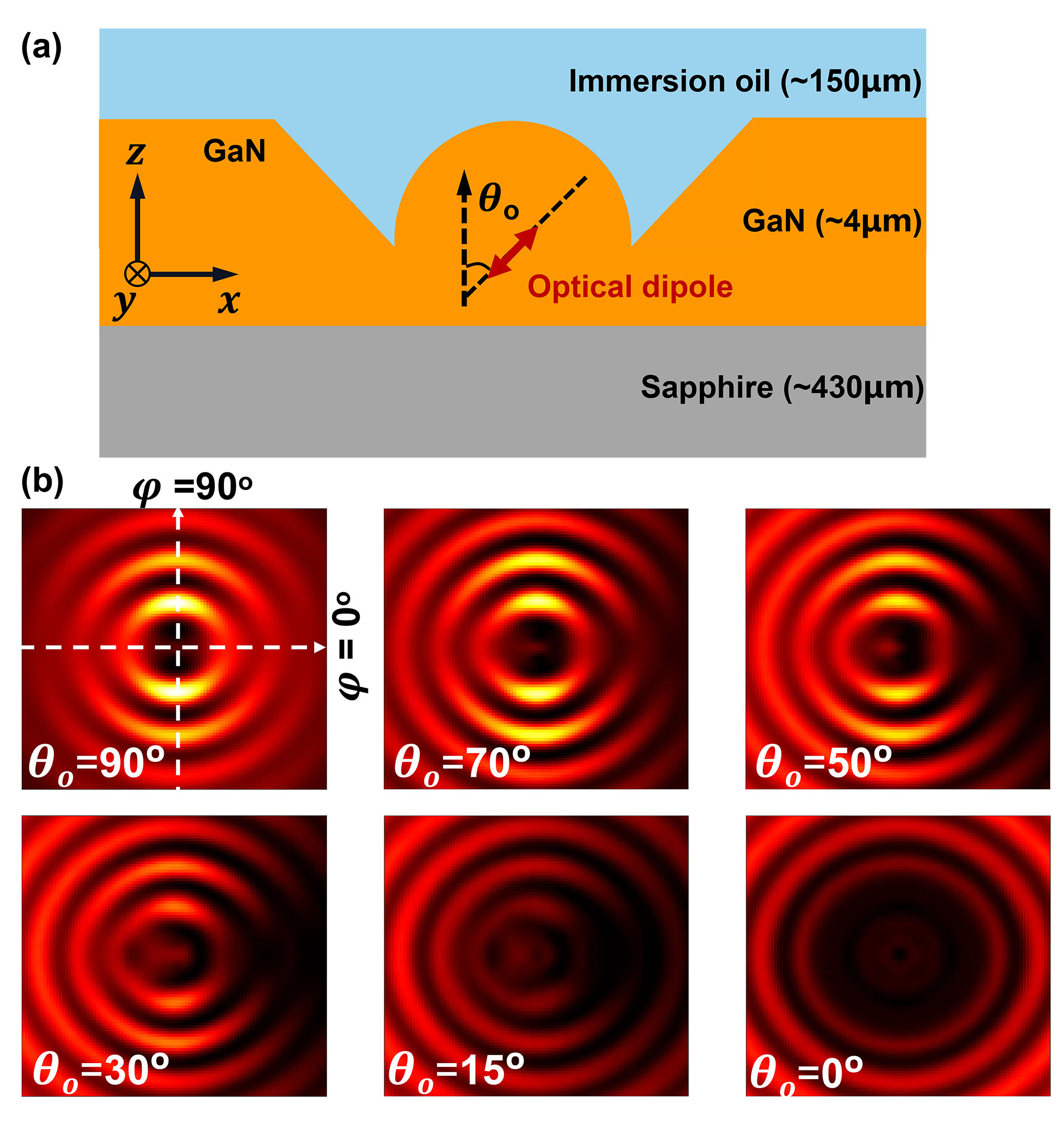}
\caption{\label{fig3} (a) The structure used in numerical (FDTD) computations. The 4 $\mu$m GaN layer is sandwiched by oil immersion layer and sapphire substrate. The diameter of the solid immersion lens is 5 $\mu$m. (b) The calculated far-field radiation patterns from the dipole are plotted for different dipole out of plane angles $\theta_{o}$. The radiation pattern has $C_{2v}$ symmetry for $\theta_{o}=90^{\circ}$, $O(2)$ symmetry for $\theta_{o}=0^{\circ}$, and $C_{1v}$ symmetry for all angles in between. Since the radiated power in the vertical direction is also a function of the dipole angle, the contrast is adjusted in the figures shown for better visibility as follows: ($\theta_{o}=90^{\circ}$, contrast 1X), ($70^{\circ}$, 1X), ($50^{\circ}$, 2X), ($30^{\circ}$, 3X), ($15^{\circ}$, 5X), and ($0^{\circ}$, 10X).}
\end{figure}

The determination of the optical dipole angles $\theta_{o}$ from the defocused images requires analytical or computational modeling. We use the finite-difference time-domain (FDTD) technique to compute the radiation patterns for different dipole angles $\theta_{o}$ and compare them with measurements. Fig.~\ref{fig3}(a) shows the model structure used in computations. A 4 $\mu$m thick GaN layer (refractive index $n=2.37$) is sandwiched between a 430 $\mu$m thick sapphire substrate ($n=1.76$) and a 150 $\mu$m thick immersion oil layer ($n=1.51$). The solid immersion lens is a hemisphere of diameter 5 $\mu$m. An optical dipole is placed in the center of the solid immersion lens. The angle $\theta_{o}$ of the dipole is varied and radiation patterns in the far field from the dipole are calculated.

Fig.~\ref{fig3}(b) shows the computed far-field radiation patterns for different representative angles $\theta_{o}$ assuming that $\varphi_{o}=0^{\circ}$ and the dipole emission wavelength is 680 nm. When $\theta_{o}=90^{\circ}$, and the dipole is oriented perpendicular to the c-axis, the radiation pattern has two-fold rotation symmetry (group $C_{2v}$) with respect to the c-axis with reflection planes at $\varphi=0^{\circ}$ and $\varphi=90^{\circ}$. When $\theta_{o}=0^{\circ}$, and the dipole is oriented along the c-axis, the radiation pattern has complete rotational symmetry (group $O$(2)). However, the power radiated in the upward direction is extremely small in this case. For all other values of $\theta_{o}$, the radiation pattern has lower symmetry (group $C_{1v}$) with a single reflection plane at $\varphi=0^{\circ}$. As shown in Fig.~\ref{fig3}(b), the lowering of symmetry in the radiation pattern from $C_{2v}$ to $C_{1v}$,  as the dipole angle is reduced from $\theta_{o}=90^{\circ}$, makes the pattern more asymmetric with respect to the $\varphi=90^{\circ}$ plane. The accuracy with which the dipole angle $\theta_{o}$ can be determined from the measured radiation pattern depends on the exact and detailed comparison between data and theory and also on the signal to noise ratio in the measured data. As shown in Figs.~\ref{fig2}(a)-(e), the measured radiation patterns of all SPEs have approximately two-fold $C_{2v}$ symmetry, which implies all SPEs are oriented perpendicular (or almost perpendicular) to the c-axis (i.e. $\theta_{o}=90^{\circ}$). However, noise in our measurements can be used to put an error margin on the value of $\theta_{o}$. By comparing the measurements with the computed images, and taking into account the noise in our data, we estimate that $\theta_{o}$ for all SPEs satisfies $70^{\circ} < \theta_{o} < 110^{\circ}$ irrespective of the wavelength and the in-plane orientation of the SPE dipole. This range specifies the noise-limited error margin in our determination of $\theta_{o}$.

Other factors that can affect the determination of the dipole angle $\theta_{o}$ include: i) the emission wavelength of a SPE can be different from 680 nm, the wavelength used in our computations, and ii) in actual samples SPEs are not exactly located at the location assumed in the computations (see Fig.~\ref{fig3}(a)). Points whose distance $r$ from a dipole satisfy $r >> \lambda/(2\pi n) \approx 50$ nm, where $n$ is the refractive index of GaN, constitute the radiation far-field region. If a dipole is away from all interfaces by at least 1 $\mu m$ ($\sim$20 times $\lambda/(2\pi n)$), and the lens structure does not favor the formation of strong standing waves, one can expect that the recorded radiation pattern will not get much affected by the exact location of the dipole inside the lens. The SPEs studied in this work were selected such that they were at least a micron away from the top and bottom interfaces of the GaN epitaxial layer to avoid surface contamination and interface defects. Furthermore, the offsets between the centers of the solid immersion lenses and the SPEs due to fabrication errors were generally less than 1 $\mu m$. This ensured that all interfaces were in the far-field regions of the SPEs. Numerical computations indeed show when the above conditions are fulfilled, the effect of the location of the dipole on the radiation pattern and on the accuracy with which $\theta_{o}$ can be determined is small enough to be almost negligible compared to the accuracy limitation imposed by the presence of noise, as discussed above. This can be seen in our experimental data as well (Figs.~\ref{fig2}(a)-(e)) where the approximate $C_{2v}$ symmetry is seen for all SPEs irrespective of their exact location within the lens. Similarly, numerical computations show that the effect of the variation in the emission wavelength of a SPE in the 600-700 nm range on the radiation pattern is small enough to be ignored.

\begin{figure}
\includegraphics[width=0.6\textwidth]{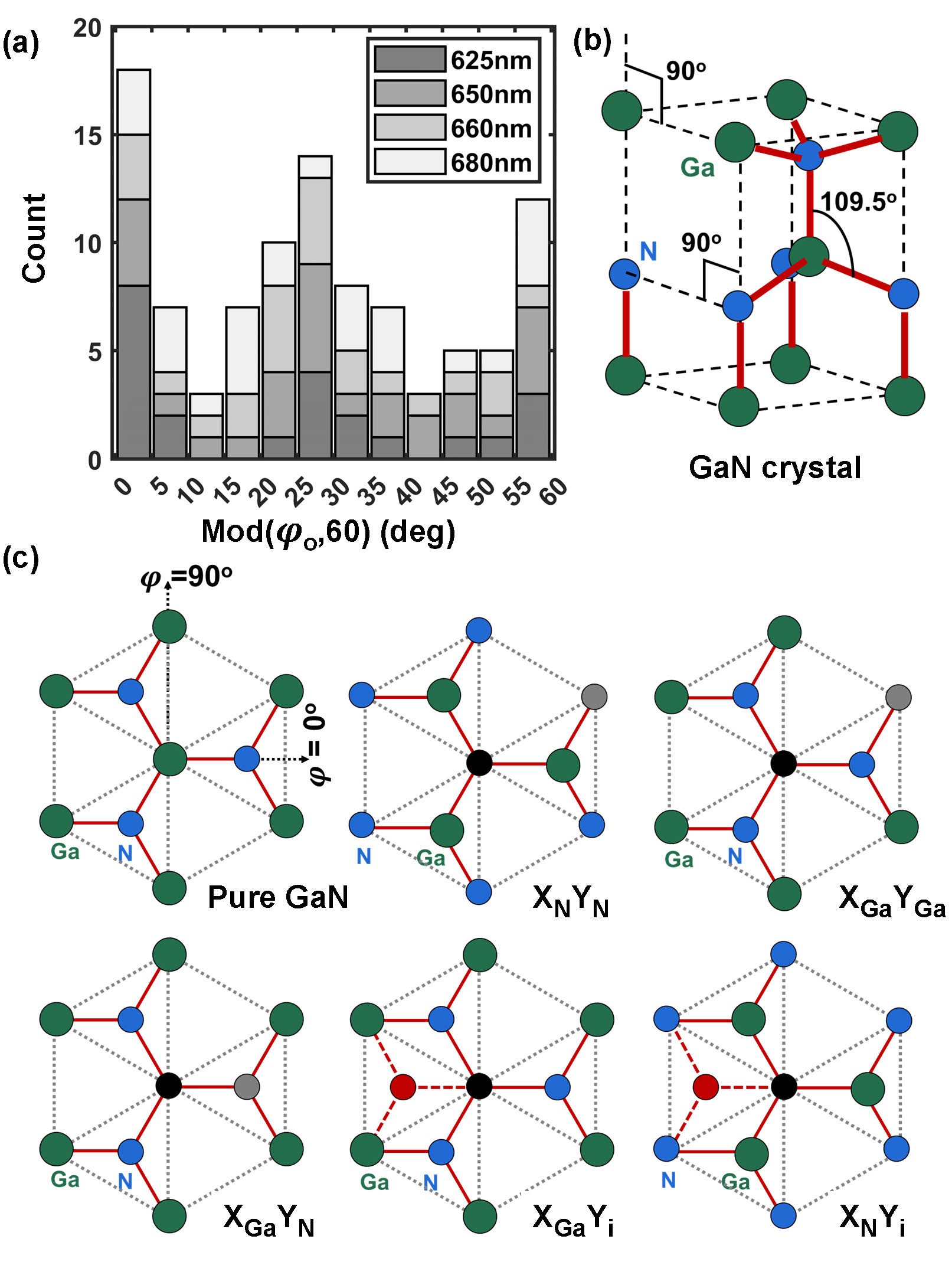}
\caption{\label{fig4} (a) The distributions of in-plane optical dipole angles $mod(\varphi_{o},60^{\circ})$ for different wavelength SPEs are shown. (b) GaN wurtzite crystal structure (side view). (c) Impurity/vacancy complexes of the form $X_{N}Y_{N}$, $X_{Ga}Y_{Ga}$, $X_{Ga}Y_{N}$, $X_{Ga}Y_{i}$, and $X_{N}Y_{i}$ are depicted. $X$,$Y$ stand for an impurity atom or a vacancy (if on a lattice site) and the subscript $i$ stands for an interstitial (shown in red color). Substitutional impurities and vacancies are shown in black or gray colors. The leftmost figure in the top row shows a pure GaN crystal (top view).}
\end{figure}

Finally, we investigate the distribution of the in-plane angle $\varphi_{o}$ of the SPEs and examine if any correlations exist between these angles and the underlying crystal structure. Since the wurtzite crystal structure has an in-plane 6-fold rotation symmetry, we classify all measured SPEs according to the angle $mod(\varphi_{o},60^{\circ})$. The resulting distribution is plotted in Fig.~\ref{fig4}(a) for close to 100 emitters with different emission wavelengths in the 600-700 nm range. The four wavelengths shown are the center wavelengths of the bins in which SPEs with emission wavelengths close to the bin center wavelength (within $\pm$5 nm) were placed in making the plot. Several interesting features are visible in this plot. First, SPEs of all wavelengths exhibit very similar angular distributions. Second, the angular distributions are fairly wide. Third, the angular distributions show two distinct peaks at angles close to $0^{\circ}$ (or $60^{\circ}$) and $30^{\circ}$. These angles correspond to the directions between the nearest Ga-N bonds and the nearest Ga-Ga (or N-N) directions in the wurtzite crystal structure of GaN, as depicted in the first sub-figure in Fig.~\ref{fig4}(c).      

The information obtained via defocused imaging about the optical dipole orientation of GaN SPEs in the 600-700 nm wavelength range can help identify the nature and origin of the SPEs. First, a single substitutional impurity or a vacancy at either the Ga or N site is expected to have $C_{3v}$ point group symmetry with the c-axis as the 3-fold axis of rotation. $C_{3v}$ group has two one-dimensional representations, $A_{1}$ and $A_{2}$, and a one two-dimensional representation $E$. Optical transitions are allowed between states with the same $A_{1}$, $A_{2}$ or $E$ symmetry, and between states with $A_{1}$ or $A_{2}$ symmetry and $E$ symmetry. In all these cases, linearly polarized light emission with a single optical dipole axis, as observed in our measurements, is not possible (unless the double degeneracy between the states with $E$ symmetry is somehow lifted). Next, we consider impurity/vacancy complexes of the form $X_{N}Y_{N}$, $X_{Ga}Y_{Ga}$, $X_{Ga}Y_{N}$, $X_{Ga}Y_{i}$, and $X_{N}Y_{i}$ (see Fig.~\ref{fig4}(c)). $X$,$Y$ stand for an impurity/substitutional atom or a vacancy (if on a lattice site) and the subscript $i$ stands for an interstitial. If one assumes that the optical dipole moment is along the axis of the $X\text{-}Y$ complex, then this family of defects could explain the measured dipole orientations (angles $\theta_{o}$ and $\varphi_{o}$) of the SPEs (note the $109.5^{\circ}$ and $90^{\circ}$ angles between the c-axis and the nearest Ga-N bonds and the nearest Ga-Ga (or N-N) directions, respectively, in Fig.~\ref{fig4}(b)). Several first principles studies have been reported for point defects in GaN~\cite{Bellotti17a,Bellotti17b,Walle19,Walle21,Bockowski21,yuan2023gan}. Point defects in which one of $X$ and $Y$ is carbon or iron and the other is carbon, hydrogen, oxygen, or a vacancy are promising candidates as computations have shown these point defects to have relatively small formation energies and their computed ZPL photon emission energies are in the neighborhood of the observed emission energies~\cite{Bellotti17a,Bellotti17b,Walle19,Walle21,Bockowski21}. However, computed optical dipole orientations have not been reported in the literature all for these defects. In the case when both $X$ and $Y$ are carbon atoms, the optical dipoles are known to be oriented along the $X\text{-}Y$ axis~\cite{Bellotti17a}. The experimentally observed departure from exact $0^{\circ}$ and $30^{\circ}$ values of $mod(\varphi_{o},60^{\circ})$ could be due to the fact that the presence of impurity atoms causes lattice distortion and, therefore, the actual positions of the atoms are not expected to be as shown in Fig.~\ref{fig4}(c)~\cite{Bellotti17a,Bellotti17b,Bockowski21}. Supporting evidence for these defects also comes from secondary ion mass spectrometry (SIMS) data. The semi-insulating GaN samples used in this work were iron doped and the iron concentration measured by SIMS was $\sim 7 \times 10^{17}$ /cm$^{3}$). SIMS also showed carbon, hydrogen, and oxygen concentrations of $\sim 5 \times 10^{16}$ /cm$^{3}$, $\sim 3 \times 10^{17}$ /cm$^{3}$, and $\sim 10^{16}$ /cm$^{3}$, respectively. Whereas the defects mentioned above are all extrinsic, recently an intrinsic antisite-vacancy complex $N_{Ga}V_{N}$ (also of the type $X_{Ga}Y_{N}$) has also been suggested as a possible candidate for SPEs with emission wavelengths close to 625 nm~\cite{yuan2023gan}. As the final word, we note here that the SPE density in our samples is less than $10^{10}$ /cm$^{3}$. This means that impurity atoms in concentrations much less than the SIMS detection limit could also underlie the observed SPEs in our samples.

\subsection{Conclusions}

In conclusion, we used the defocused imaging technique to study the optical dipole structures and orientations of defect SPEs in GaN. Our experimental results, backed up by FDTD calculations, show that GaN defect SPEs in the 600-700 nm wavelength range consist of a single optical dipole and the dipole orientation is approximately perpendicular to the crystal c-axis. The angular orientation of the dipole moment in the plane perpendicular to the c-axis exhibits a distribution that shows peaks centered at the angles corresponding to the nearest Ga-N bonds and also at the angles corresponding to the nearest Ga-Ga (or N-N) directions. Our findings are consistent with these SPEs being point defects. We hope that this work will stimulate further theoretical and experimental studies on the nature and origin of GaN SPEs.

\begin{acknowledgement}

The authors thank Len van Deurzen for SIMS information of the sample. This work was supported by the Cornell Center for Materials Research with funding from the NSF MRSEC program (DMR-1719875) and also by the NSF-RAISE:TAQS (ECCS-1838976).

\end{acknowledgement}



\bibliography{ref}

\end{document}